\documentclass[12pt]{article}
\usepackage[english,german,french,polish]{babel}
\usepackage[T1]{fontenc}
\usepackage{amsfonts}

\begin{document}

\selectlanguage{english}

\baselineskip 0.73cm
\topmargin -0.4in
\oddsidemargin -0.1in

\let\ni=\noindent

\renewcommand{\thefootnote}{\fnsymbol{footnote}}

\newcommand{\SM}{Standard Model }

\pagestyle {plain}

\setcounter{page}{1}



~~~~~~
\pagestyle{empty}

\begin{flushright}
IFT-- 08/5
\end{flushright}

\vspace{0.4cm}

{\large\centerline{\bf A Hidden Valley model of cold dark matter}}

{\large\centerline{\bf with photonic portal}}

\vspace{0.5cm}

{\centerline {\sc Wojciech Kr\'{o}likowski}}

\vspace{0.3cm}

{\centerline {\it Institute of Theoretical Physics, University of Warsaw}}

{\centerline {\it Ho\.{z}a 69, 00--681 Warszawa, ~Poland}}

\vspace{0.6cm}

{\centerline{\bf Abstract}}

\vspace{0.2cm}

\begin{small}

\begin{quotation}

In the discussed model, the cold dark matter consists of Dirac spin-1/2 fermions, sterile from all \SM charges,
where masses are generated by a nonzero vacuum expectation value of a field of scalar bosons, also assumed to be 
sterile. For convenience, these sterile particles have beeen called {\it sterinos} and {\it sterons}, 
respectively. It has been conjectured that our sterile world of sterinos and sterons can communicate with the
familiar \SM world not only through gravity, but also through a photonic portal provided by a very weak effective interaction involving the electromagnetic field $F_{\mu\,\nu} = \partial_\mu A_\nu - \partial_\nu 
A_\mu $ coupled to the sterino and steron fields. 
 
\vspace{0.6cm}

\ni PACS numbers: 14.80.-j , 04.50.+h , 95.35.+d 

\vspace{0.3cm}

\ni April 2008

\end{quotation}
 
\end{small}

\vfill\eject

\pagestyle {plain}

\setcounter{page}{1}

\vspace{0.3cm}

\ni {\bf 1. Introduction}

\vspace{0.3cm} 

In a recent paper [1] we have proposed a simple model for cold dark matter  consisting of spin-1/2 fermions, sterile from all \SM charges, where masses are generated by a nonzero vacuum expectation value of a field of scalar bosons, also assumed to be sterile. We have conjectured that, after the electroweak symmetry is broken, the fields of these sterile particles, $\psi$ and $\varphi$ respectively, interact with the electromagnetic field $F_{\mu\,\nu} = \partial_\mu A_\nu - \partial_\nu A_\mu $ through a photonic portal provided by the very weak effective interaction

\vspace{-0,1cm}

\begin{equation}
-\frac{f}{M^2}\,F_{\mu\,\nu}\,\varphi F^{\mu\,\nu}\,\varphi - \frac{f'}{M^2}\, (\bar\psi \sigma_{\mu\,\nu} \psi )\, F^{\mu\,\nu}\varphi \;,
\end{equation}

\ni where $M$ is a very large mass scale, while $f$ and $f'$ are unknown dimensionless coupling constants. Besides, $\psi$ and $\varphi$ fields interact gravitationally, as well as participate in the mass-generating 
Higgs-type coupling

\begin{equation}
-\bar{\psi}\,Y\, \psi\, \varphi + \frac{1}{2}\mu^2 \varphi^2 - \frac{1}{4}\lambda \varphi^4\;\;\;{\rm with}\;\;\; <\!\!\varphi\!\!>_{\rm vac}\, \neq 0 
\end{equation}

\vspace{-0.2cm}

\ni (if there are more generations of $\psi $ fermions, then $Y$ is a constant matrix). Here, the physical scalar field is identified with $\varphi_{\rm ph} \equiv \varphi\, - <\!\!\varphi\!\!>_{\rm vac}$. We will assume tentatively that $\psi $ particles are Dirac fermions. Note that the third term $-{f''}/{M^2}\, (\bar\psi \sigma_{\mu\,\nu} \psi)(\bar\psi \sigma^{\mu\,\nu} \psi)$ might be introduced into the effective interaction (1), suggesting then a kind of sterile universality, where it would be natural to put $f : f' : f'' = 1 : 2 : 1$. For convenience, we have called $\psi $ and $\varphi $ sterile particles {\it sterinos} and {\it sterons}, respectively{\footnote{A natural alternative for  the photonic portal to the sterile world my be a Higgs portal provided by a very weak direct interaction between \SM Higgs bosons and some sterile scalars (called in our case sterons) causing a mixing of both ({\it cf.} Ref. [2] for a recent discussion including the Sommerfeld corrections to this interaction).}}.

It is not difficult to see that the effective interaction (1), when added to the electromagnetic Lagrangian $-(1/4)\,F_{\mu\,\nu}\, F^{\mu\,\nu} - j_\mu A^\mu $, leads to the following electromagnetic field equation:

\vspace{-0.2cm}

\begin{equation}
\partial_\nu F^{\mu\,\nu} = -\left(j^\mu + \delta j^\mu \right) \,,
\end{equation}

\vspace{-0.1cm}

\ni where the additional current

\vspace{-0.2cm}

\begin{equation}
\delta j^\mu \equiv \frac{4}{M^2} \partial_\nu\left[\varphi \left(f\,\varphi F^{\mu\,\nu}   + \frac{1}{2}f'\, \bar\psi \sigma^{\mu\,\nu} \psi \right)\right] 
\end{equation}

\vspace{-0.1cm}

\ni is a quasi-magnetic correction to the \SM electromagnetic current $j^\mu$, induced by the photonic effective interaction (1). For this correction we get $\partial_\mu j^\mu \equiv 0$ identically, while the local conservation equation $\partial_\mu j^\mu = 0$ holds dynamically. So, the effective interaction (1) providing our photonic portal is a quasi-magnetic interaction.

The particle models, where beside the \SM sector there exists a sterile sector interacting through new forces with itself as well as with the \SM sector, have been called Hidden Valley models [3]. Our model of sterinos and sterons interacting very weakly through the photonic portal provided by a new quasi-magnetic force is a natural specific realization within such a class of models. In this option, photons are common elements which link (very weakly) both sectors: sterile Hidden Valley and active Standard Model. This happens, of course, after the electroweak symmetry is broken and photons emerge.

Now, let us consider the simple annihilation and decay channels for sterons and sterinos, and present the formulae for corresponding cross-sections and rates.

The simplest annihilation channel for a physical steron pair and decay channel for a single physical steron are

\begin{equation} 
({\rm steron})({\rm steron}) \rightarrow \gamma\, \gamma
\end{equation}

\ni and

\begin{equation} 
({\rm steron}) \rightarrow \gamma\, \gamma\,,
\end{equation}

\ni respectively. From the interaction (1) we can derive respectively the following formulae for the corresponding total cross section (multiplied by the relative velocity) and total rate [1]:

\begin{equation}
\sigma({\rm stn\; stn}\rightarrow{\gamma \gamma})  2v_{\rm stn} = \frac{1}{\pi} \left( \frac{f}{M^2} \right)^{\!\!2} \omega^2_{\rm stn} 
\end{equation}

\ni in the steron pair centre-of-mass frame, and  

\begin{equation}
\Gamma({\rm stn}\rightarrow{\gamma \gamma}) = \frac{1}{8\pi} \left(
\frac{f<\!\!\varphi\!\!>_{\rm vac} }{M^2}\right)^2\,\omega^3_{\rm stn}
\end{equation}

\ni at rest of steron: $\omega_{\rm stn} = m_{\rm stn}$, where  $\omega_{\rm stn} \equiv \sqrt{\vec{p}^{\,2}_{\rm stn}  +m^2_{\rm stn}}$ and $v_{\rm stn} \equiv |\vec{p}_{\rm stn}| /\omega_{\rm stn} $ are the steron energy and velocity.

The more complicated annihilation channel

\begin{equation} 
({\rm steron})({\rm steron}) \rightarrow e^+e^-\, \gamma
\end{equation}

\ni and decay channel

\begin{equation} 
({\rm steron}) \rightarrow e^+e^-\, \gamma
\end{equation}

\ni get respectively the total cross-section [1]

\begin{equation} 
\sigma({\rm stn\; stn}\rightarrow{e^+ e^- \gamma})\, 2v_{\rm stn} =\frac{4}{(3\pi)^3} \left( \frac{e\,f}{M^2} \right)^2 \omega^2_{\rm stn} 
\end{equation}

\ni in the centre-of-mass frame,  and the total rate [1]

\vspace{-0.2cm}

\begin{equation}
\Gamma({\rm stn}\rightarrow{e^+ e^- \gamma}) = \frac{1}{6\pi^3} \left(\frac{e\,f <\!\!\varphi\!\!>_{\rm vac} }{M^2}\right)^{\!\!2} \,\omega^3_{\rm stn} 
\end{equation}

\vspace{0.2cm}

\ni at rest: $\omega_{\rm stn} = m_{\rm stn}$, if the electron mass $m_e$ can be neglected. Here, the interaction (1) is used together with the \SM electromagnetic coupling $ e \bar{\psi}_e \gamma^\mu \psi_e A_\mu $ for electrons ($e = |e|$).

The simplest annihilation channel for a sterino-antisterino pair
 
\begin{equation}
({\rm antisterino})({\rm sterino}) \rightarrow \gamma\, ({\rm steron})  
\end{equation}

\ni requires, due to the interaction (1), the total cross-section [1]

\begin{equation}
\sigma({\rm asto\,sto} \rightarrow \gamma \;{\rm stn}) \,2v_{\rm sto} = \frac{2} {3\pi } \left(\frac{f'}{M^2} \right)^{\!\!2} \left(E^2_{\rm sto} + 2m^2_{\rm sto}\right)\left(1 - \frac{m^2_{\rm stn}}{4E^2_{\rm sto}} \right)
\end{equation}

\vspace{0.2cm}

\ni in the sterino--antisterino centre-of-mass frame, where  $ E_{\rm sto} \equiv \sqrt{\vec{p}^{\,2}_{\rm sto} + m^2_{\rm sto}}$ and $v_{\rm sto}  \equiv |\vec{p}_{\rm sto} |/E_{\rm sto}$ are the sterino energy and velocity. Note that in contrast to one-physical-steron states, one-sterino states are stable under the interaction (1) of our photonic portal 

The more complicated annihilation channel
 
\begin{equation}
({\rm antisterino})({\rm sterino}) \rightarrow e^+e^- 
\end{equation}


\ni has the total cross-section [1]

\begin{equation}
\sigma({\rm asto\,sto} \rightarrow e^+ e^-)\, 2v_{\rm sto} = \frac{4} {3\pi }\,\left(\frac{e f' <\!\!\varphi\!\!>_{\rm vac}} { M^2} \right)^{\!\!2}\, \frac{E^2_{\rm sto}+2m^2_{\rm sto}}{E^2_{\rm sto}} 
\end{equation}

\vspace{0.2cm}

\ni in the sterino-antisterino centre-of-mass frame, if the electron mass is negligible. Here, the interaction (1) is applied together with the \SM electromagnetic coupling $e \bar{\psi}_e\gamma^\mu{\psi}_e A_\mu $ for electrons.

The elastic scattering of electrons on sterinos,


\begin{equation}
e^- ({\rm sterino})\rightarrow e^- ({\rm sterino})\,, 
\end{equation}

\vspace{-0.1cm}

\ni gets the differential cross-section [1]


\begin{equation}
\frac{d \sigma(e^- {\rm sto} \rightarrow e^- {\rm sto})}{d \Omega_e} = \left( \frac{e f' <\!\!\varphi\!\!>_{\rm vac}} {\pi M^2}\right)^{\!\!2} \frac{1 + \sin^2\frac{\theta_e}{2}}{\sin^2\frac{\theta_e}{2} } 
\end{equation}

\vspace{0.2cm}

\ni in the sterino rest frame: $E_{\rm sto} = m_{\rm sto}$, if the sterino recoils are neglected: $E'_{\rm sto} \simeq E_{\rm sto} = m_{\rm sto}$ {\it i.e.}, $E_e \ll m_{\rm sto}$. Here, the interaction (1) is used together with the \SM electromagnetic coupling $e \bar{\psi}_e\gamma^\mu{\psi}_e A_\mu $ for electrons. The forward singularity in Eq. (18) is softer than in the corresponding \SM electron differential cross-section on, say, point-like protons, valid with $E_e \ll m_p$ (Mott cross-section [4]).

The same formula (18) as for electrons scattered elastically on sterinos holds {\it mutatis mutandis} for elastic
scattering of point-like protons on sterinos. This scattering is the simplest interaction between nuclei and cold dark matter composed of heavy sterinos, subject to possible direct detection experiments for cold dark matter [5]. Possible indirect detection experiments for cold dark matter [6] correspond to our previous formulae in this Section.

\vspace{0.3cm}

\ni {\bf 2. Sterinos and the thermal freeze-out }

\vspace{0.3cm} 

Our main question in the present paper is, what mass is required for sterinos in order that they may freeze out thermally in the early Universe, forming  the cold dark matter observed today. To answer this question a number of tentative assumptions will be required.

In the case of  thermal freeze-out processes in the early Universe, the order-of-magnitude theoretical estimation for the  relic dark-matter abundance has the form [7]


\begin{equation}
\Omega_{\rm DM}  h^2 \simeq \frac{3\times 10^{-27}\,{\rm cm}^3 {\rm s}^{-1}}{<\!\!\sigma_{\rm ann} v_{\rm DM}\!\!>}\,,
\end{equation}

\vspace{0.2cm}

\ni where $<\!\!\sigma_{\rm ann} v_{\rm {DM}\!\!>}$ denotes the thermal average of the dark-matter total annihilation cross-section multiplied by relative velocity, while  $h$ stands for the today's value of scaled Hubble parameter, $h = 72 \pm 3{\rm (stat)} \pm 7 {\rm (syst)}/100$, where $H_0 = 100 h$ km ${\rm s}^{-1}{\rm Mpc}^{-1}$ .

The WAMP experimental estimate for the relic dark-matter abundance is [7]


\begin{equation}
\Omega_{\rm DM}  h^2 \simeq 0.1 \,.
\end{equation}


\ni Thus, Eqs. (19) and (20) lead to
 
\vspace{0.1cm}

\begin{equation}
<\sigma_{\rm ann} v_{\rm {DM}>} \simeq 3\times 10^{-26} {\rm cm}^3{\rm s}^{-1} \simeq {\rm pb}\,  \simeq \frac{8}{\pi} \frac{10^{-3}}{{\rm TeV}^2}
\end{equation}

\vspace{0.3 cm}

\ni in the units where $c = 1$ and $\hbar c = 1$ (${\rm pb} = 10^{-36} {\rm cm}^2$). The thermal-equilibrium experimental value (21) happens to be consistent with the typical size of weak-interaction cross-sections, providing therefore a strong numerical argument for the weakly interacting massive particles (WIMPs) as candidates for cold dark matter (as well as for the thermal-equilibrium mechanism of their decoupling in the early Universe). Our sterinos are not introduced as WIMPs, and so their interaction strength ought to be estimated, after some magnitudes for the coupling constant $f'$ and mass scale $M$ are tentatively established.

In the case of our model of cold dark matter consisting of sterinos interacting through the photonic portal, we can put approximately

\vspace{0.1cm}

\begin{equation}
\sigma_{\rm ann} v_{\rm {DM}} \simeq \sigma({\rm asto\, sto} \rightarrow \gamma\, {\rm stn}) 2v_{\rm sto} =   \frac{2}{3\pi} \left(\frac{f'}{M^2}\right)^{\!\!2}\left(E^2_{\rm sto}+ 2m^2_{\rm sto}\right) \left(1 -   \frac{m^2_{\rm stn}}{4 E^2_{\rm sto}}\right) \,,
\end{equation}

\vspace{0.2cm}

\ni when we make use of the leading sterino-antisterino annihilation cross-section (14). We will imagine for simplicity that in the Universe there is no asymmetry between sterinos and antisterinos (no excess of either).

If $E_{\rm sto} \simeq m_{\rm sto}$ ({\it i.e.}, $|\vec{p}_{\rm sto}|/m_{\rm sto} \ll 1 $) and tentatively 

\vspace{0.1cm}

\begin{equation}
m_{\rm stn} \sim m_{\rm sto} \,,
\end{equation}

\vspace{0.1cm}

\ni then Eq. (22) gives

\vspace{0.1cm}

\begin{equation}
\sigma_{\rm ann} v_{\rm {DM}} \simeq \frac{2}{\pi}  \left(\frac{f'}{M^2}\right)^{\!\!2} m^2_{\rm sto}\left(1 -   \frac{m^2_{\rm stn}}{4m^2_{\rm sto}}\right) \sim \frac{3}{2\pi}  \left(\frac{f'}{M^2}\right)^{\!\!2} m^2_{\rm sto}\,.
\end{equation}

\vspace{0.3cm}

\ni Thus, when the thermal-equilibrium experimental value (21) {\it is accepted for sterinos}{\footnote{Then, $<\!\!\sigma_{\rm ann} v_{\rm {DM}}\!\!>_{\rm sto} \simeq {\rm pb} \simeq <\!\!\sigma_{\rm ann} v_{\rm {DM}}\!\!>_{\rm WIMP}$ as well as $(\Omega_{\rm DM} h^2)_{\rm sto} \simeq 0.1 \simeq (\Omega_{\rm DM} h^2)_{\rm WIMP}$. This implies the necessary condition $x_{f \,{\rm sto}} \simeq x_{f \,{\rm WIMP}}$ with $x_{\!f} \equiv m_{\rm DM}/T_{\!f}$, in consequence of the basic formula for the relic dark-matter abundance (less approximate than (19)): $\Omega_{\rm DM}h^2 \simeq 1.07\times 10^9 x_{\!f}\,{\rm GeV}^{-1}/(g^{1/2}_* M_{\rm Pl} <\!\!\sigma_{\rm ann} v_{\rm {DM}}\!\!> $), being valid in this form when $<\!\!\sigma_{\rm ann} v_{\rm {DM}}\!\!>$ contains approximately only  $S$ wave (as in Eq. (24)) [7]. Here, $T_{\!f}$ is the freeze-out temperature, $g_*$ denotes the total number of effectively relativistic degrees of freedom in the \SM thermal plasma at the time of freeze-out and $M_{\rm Pl} = 1.22\times 10^{19}$ GeV stands for the Planck mass. From the equation $x_{\!f} = \ln \left[0.038\, g_{\rm DM}\, M_{\rm Pl}\,m_{\rm DM}<\!\!\sigma_{\rm ann} v_{\rm {DM}}\!\!> /(g_* x_{\!f})^{1/2}\right] $ defining $x_{\!f}$ [7], we infer that $x_{\!f \,{\rm sto}}/x_{\!f \,{\rm WIMP}} - 1 + (1/2)\ln (x_{\!f \,{\rm sto}}/x_{\!f \,{\rm WIMP}})/x_{\!f \,{\rm WIMP}} \simeq \ln (g_{\rm sto} m_{\rm sto}/ g_{\rm WIMP}m_{\rm WIMP})/x_{\!f \,{\rm WIMP}}$, where $g_{\rm DM} =g_{\rm sto}$ or  $g_{\rm WIMP}$ counts internal degrees of freedom of  sterino or WIMP. Then, from $x_{\!f \,{\rm sto}}/x_{\!f \,{\rm WIMP}} \simeq 1$ we obtain the necessary condition $|\ln (g_{\rm sto} m_{\rm sto}/ g_{\rm WIMP}m_{\rm WIMP})/x_{\!f \,{\rm WIMP}}| \ll 1$ for applicability of WIMP freeze-out formula (19) to our sterinos as the cold dark matter. Here, $x_{\!f \,{\rm WIMP}} \simeq 25$. Thus, for {\it e.g.} $g_{\rm sto} m_{\rm sto}/ g_{\rm WIMP}m_{\rm WIMP} = m_{\rm sto}/m_{\rm WIMP} \sim 1$ to $\,6\,$ (taking $m_{\rm WIMP} \sim$ 100 GeV and $m_{\rm sto} \sim 0.1 $ to 0.6 TeV, the estimate (26) as the upper limit of $m_{\rm sto}$), we get $\ln (g_{\rm sto} m_{\rm sto}/ g_{\rm WIMP}m_{\rm WIMP})/ x_{\!f \,{\rm WIMP}} \sim 0$ to 0.07, what is $\ll 1 $ neatly.

~~}}, the formula (24) implies

\begin{equation}
m_{\rm stn} \sim m_{\rm sto} \sim \frac{2\times 10^{-3/2}}{\sqrt3} \frac{2M^2}{f'}\,\frac {1}{\rm TeV}  \simeq  \frac{1}{27} \frac{2M^2}{f'}\,\frac{1}{\rm TeV} \;\,,
\end{equation}

\vspace{0.3cm}

\ni since here $<\!\!\sigma_{\rm ann} v_{\rm DM}\!\!>\; \simeq \sigma_{\rm ann} v_{\rm {DM}}$ as $\sigma_{\rm ann} v_{\rm {DM}}$ depends only on $S$ wave (see Eq. (24)). From Eq. (25) we can obtain the possible mass estimation

\vspace{0.1cm}

\begin{equation}
m_{\rm stn} \sim m_{\rm sto} \sim 27\,\frac{f'}{2} \;{\rm TeV} \sim 0.6\,{\rm TeV}\,,
\end{equation}

\ni if we put tentatively

\begin{equation}
m_{\rm sto} \sim M \;\;{\rm and}\;\; f' \sim 2e^2/4 = e^2/2 = 2/43.6 \,,
\end{equation}

\vspace{0.1cm}

\ni where $e^2/4\pi =  \alpha = 1/137$ (with $f \sim e^2/4$, the estimate $f' \sim e^2/2$ would be consistent with the sterile universality $f : f' : f'' = 1 : 2 : 1$; the factor 1/4 in $f$ is suggested by the normalization of the term $-(1/4) F_{\mu \nu}F^{\mu \nu}$ in the Lagrangian).
 
We can see that --- with the sterino mass $ m_{\rm sto}$ as given in Eq. (25) or, possibly, Eq. (26) --- the thermal-equilibrium decoupling mechanism (leading to the formula (19)) can work in the case of  the
sterino-antisterino pairs annihilating according to our cross-section (22). This statement is valid under the tentative assumptions of  $m_{\rm stn} \sim m_{\rm sto}$ and, possibly, $m_{\rm sto} \sim M$ and $f' \sim e^2/2$. 

\vspace{0.4cm}

\ni {\bf 3. Conclusions and final remarks}

\vspace{0.4cm}

The picture emerging from our model (proposing sterinos and sterons as particles responsible for cold dark matter) looks as follows. Sterinos are stable under the  interaction of our photonic portal, while physical sterons appear as unstable (they turn out to be also unstable on the Universe time-scale). For sterinos, the thermal-equilibrium freeze-out mechanism can work, if their mass is $m_{\rm sto} \!\sim\! $ 0.6 TeV (with $m_{\rm stn} \!\sim\! m_{\rm sto} \!\sim\! M\!$ and $\!f' \sim e^2/2 = 1/21.8$). Thus, along the thermal option, sterinos interacting through the photonic portal may find a place among candidates for the observed cold dark matter. 

The strength of our quasi-magnetic interaction of sterinos, defined in Eq. (1) as $f'/M^2$, is $\sim e^2/{2M^2} \sim 10^{-1}/$TeV$^2 \sim 10^{-7}/$GeV$^2$, while for weak interactions the strength is given by the Fermi constant $G_F/\sqrt2 \equiv e^2/(8M^2_W \sin^2\theta_W) = 1.1664\times 10^{-5}/(\sqrt2$ GeV$^2)$~~ $(\hbar c = 1$) [10]. So, the ratio of both turns out to be $\sim 10^{-2}$.

Note finally that the interaction of sterino-antisterino pairs with photons following from Eq. (1), where $\varphi \equiv <\!\!\varphi\!\!>_{\rm vac} + \varphi_{\rm ph}$, gets the effective magnetic-like form

\begin{equation}
\frac{f'<\!\!\varphi\!\!>_{\rm vac}}{M^2)}\left(\bar\psi \sigma_{\mu \nu} \psi \right) F^{\mu \nu} \sim \frac{e^2<\!\!\varphi\!\!>_{\rm vac}}{2M^2)} \left(\bar\psi \sigma_{\mu \nu} \psi \right) F^{\mu \nu} \;, 
\end{equation}

\ni proportional to the sterino effective magnetic-like moment~$\mu_{\rm eff} \,\equiv\, e^2<\!\!\varphi\!\!>_
{\rm vac}\!/\!(2M^2) \!\sim\! 10^{-1}\!\!<\!\!\varphi\!\!>_{\rm vac}\!/$TeV$^2 \!\sim\! 10^{-7}\!\!<\!\!\varphi\!\!>_{\rm vac}\!/$GeV$^2$.

If behind the effective interaction (1) there stood a new mediating antisymmetric tensor field $A^{\mu \nu}$ (of dimension one) with the mass $M$, coupled to the total "tensor current"\, $F_{\mu \nu} \varphi + \bar\psi \sigma_{\mu\,\nu} \psi$, then the sterile universality $f : f' : f'' = 1 : 2 : 1$ would be realized in effective interactions of the photonic portal. In this case, the following extended electromagnetic Lagrangian would work:

\begin{equation}
{\cal{ L }} = -\frac{1}{4}F_{\mu\,\nu} F^{\mu\,\nu} - j_\mu A^\mu +\frac{1}{4} \left[\left(\partial_\lambda A_{\mu \nu}\right)\left(\partial^\lambda A^{\mu \nu}\right) - M^2 A_{\mu \nu}A^{\mu \nu}\right] - \sqrt{f}\left(F_{\mu \nu}\varphi + \bar\psi \sigma_{\mu\,\nu} \psi \right) A^{\mu \nu} \;, 
\end{equation}

\ni giving two independent field equations for $A^\mu $ and $A^{\mu \nu}$:

\begin{equation}
\partial_\nu \left( F^{\mu\,\nu} + 2\sqrt{f} \varphi A^{\mu \nu} \right) = - j^\mu \;\;,\;\; \left( \Box - M^2\right) A^{\mu \nu} = - 2\sqrt{f} \left(F^{\mu \nu}\varphi + \bar\psi \sigma^{\mu \nu} \psi \right) \;, 
\end{equation}

\ni with $F_{\mu \nu} = \partial _\mu A_\nu - \partial_\nu A_\mu $ and $\partial_\nu \partial^\nu  = -\Box $). The first Eq. (30) has the form of $\Box A^\mu = -(j^\mu + \delta j^\mu)$, where $\delta j^\mu\equiv  2\sqrt{f}\partial_\nu( \varphi A^{\mu \nu})$, thus $ \partial_\mu \delta j^\mu \equiv 0$ identically, showing that $\delta j^\mu $ is a quasi-magnetic correction to the \SM electromagnetic current (so $\delta j^\mu $ carries no gauge charge). If the momentum transfers through the field $A^{\mu \nu}$ are approximately neglected {\it i.e.}, $A^{\mu \nu} \simeq (2\sqrt{f}/M^2)(F^{\mu \nu}\varphi + \bar\psi \sigma^{\mu \nu} \psi)$ from the second Eq. (30), then $\delta j^\mu \simeq (4f/M^2)\partial_\nu[\varphi (F^{\mu \nu}\varphi + \bar\psi \sigma^{\mu \nu} \psi)]$ (see Eq. (4)). The first Eq. (30) can be also rewritten in the form of $\partial_\nu(F^{\mu \nu} + \delta  F^{\mu \nu}) = -j^\mu$, where $\delta  F^{\mu \nu} \equiv 2\sqrt{f} \varphi A^{\mu \nu}$ is a quasi-magnetic correction to the \SM electromagnetic field $F^{\mu \nu}$, giving $\partial_\nu \delta F^{\mu \nu} \equiv \delta j^\mu$. If $A^{\mu \nu} \simeq (2\sqrt{f}/M^2)(F^{\mu \nu}\varphi + \bar\psi \sigma^{\mu \nu} \psi)$, then $\delta F^{\mu \nu} \simeq (4f/M^2)\varphi (F^{\mu \nu}\varphi + \bar\psi \sigma^{\mu \nu} \psi)$. However, if we wanted to make this substitution for $A^{\mu \nu}$  in the interaction term in Eq. (29), we should multiply that term by 1/2 because of the square operation. In the case of our tentative assumption $f \sim e^2/4$, the coupling constant in the new universal interaction of $\varphi, \psi$ and $F_{\mu \nu}$ with $A^{\mu \nu}$ in Eqs. (30) would be $ 
2\sqrt{f} \sim e$ (it might be even = $e$).

It is still possible, however,  that --- in reality --- there is no direct interaction portal to the sterile world. In such a puristic option the sterile world can communicate with the \SM world only through gravity. Then, only gravitons and, perhaps, also dilaton-like scalar and/or axion-like pseudoscalar partners of gravitons [8, 9] can mediate between both worlds as well as within the sterile world itself. Such an "isolated"\, sterile world still may be responsible for the fundamental phenomenon of cold dark matter (which, in this case, is not detectable "directly"\,$\!$ nor "indirectly"\,$\!$  in the technical sense used previously in this paper). Perhaps, spin-0 partners of gravitons may provide a gravitational interpretation of the equally fundamental phenomenon of dark energy.
 
\vfill\eject

\vspace{0.4cm}

{\centerline{\bf References}}

\vspace{0.4cm}

\baselineskip 0.73cm

{\everypar={\hangindent=0.65truecm}
\parindent=0pt\frenchspacing

{\everypar={\hangindent=0.65truecm}
\parindent=0pt\frenchspacing

~[1]~W.~Kr\'{o}likowski, arXiv: 0712.0505 [{\tt hep--ph}].

\vspace{0.2cm}

~[2]~J. March-Russell, S.M. West, D. Cumberbath and D.~Hooper, arXiv: 0801.3440v2 [{\tt hep-ph}].

\vspace{0.2cm}

~[3]~M.J. Strassler  and K.M. Zurek,  {\it Phys. Lett.} {\bf B 651}, 374 (2007); {\it cf} also J.~Kummer and J.~Wells, {\it Phys. Rev.} {\bf D 74}, 115017 (2006); T.~Han, Z.~Si, K.M.~Zurek and M.J.~Strassler, arXiv: 0712.2041 [{\tt hep-ph}].

\vspace{0.2cm}

~[4]~J.D.~Bjorken and S.D.~Drell, {\it Relativistic Quantum Mechanics} (McGraw-Hill, New York, 1964). 

\vspace{0.2cm}

~[5]~For a recent review {\it cf.} R.~Essig, arXiv: 0710.1668 [{\tt hep-ph}].

\vspace{0.2cm}
 
~[6]~For a recent review {\it cf.} D.~Hooper, arXiv: 0710.2062 [{\tt hep-ph}]. 

\vspace{0.2cm}

~[7]~For recent reviews {\it cf.} G. Bartone, D.~Hooper and J.~Silk, {\it Phys. Rept.} {\bf 405}, 279 (2005); M.~Taoso, G.~Bartone and A.~Masiero, arXiv: 0711.4996 [{\tt astro-ph}]; {\it cf.} also E.W.~Kolb and S.~Turner, {\it Early Universe} (Addison-Wesley, Reading, Mass., 1994); and K.~Griest and D.~Seckel, {\it Phys. Rev.} {\bf D 43}, 3191 (1991). 

\vspace{0.2cm}

~[8]~P. Jordan, {\it Schwerkraft und Weltall} (Vieweg, Braunschweig,1955); M.~Fierz, {\it Helv. Phys. Acta} {\bf 29}, 128 (1956); C.~Brans and R.H.~Dicke, {\it Phys. Rev.} {\bf 124}, 925 (1961); T.~Damour and K.~Nordtvadt, {\it Phys. Rev. Lett.} {\bf 70}, 2217 (1993); R.~Catena, M.~Pietroni  and L.~Scarabello,
 {\tt astro--ph/0604492}.

\vspace{0.2cm}

~[9]~X.~Calmet, arXiv: 0708.2767 [{\tt hep-ph}]; R.~Catena and J.~M\"{o}ller, arXiv: 0709.1931 [{\tt hep-ph}]. 

\vspace{0.2cm}

[10]~Particle Data Group, {\it Review of Particle Physics, J. Phys}, {\bf G 33}, 1 (2006).

\vfill\eject

\end{document}